\documentclass[a4paper,11pt]{article}
\usepackage{jcappub}
\usepackage{lineno}

\arxivnumber{2604.12521} 

\title{POLARIS: A Sparse Radial Neutrino Telescope Design for the Pacific Ocean}

\author[a]{Karolin Hymon,}
\author[a,b]{Alexander Chen,}
\author[a,c]{Meng-Xue (Mark) Tsai,}
\author[a]{Wan-Ting Hseu,}
\author[a,d]{Tzu-Hsuan (Shane) Su,}
\author[a]{and Anatoli Fedynitch}

\affiliation[a]{Institute of Physics, Academia Sinica, Taipei City, 11529, Taiwan}
\affiliation[b]{New York University, New York, NY 10003, USA}
\affiliation[c]{National Central University, Zhongli 320317, Taiwan}
\affiliation[d]{University of Illinois at Urbana-Champaign, Urbana, IL 61801, USA}

\emailAdd{khymon@as.edu.tw}
\emailAdd{anatoli@as.edu.tw}

\abstract{The cubic-kilometer neutrino telescopes have opened neutrino astronomy as an observational discipline. The recent detection of KM3-230213A, the highest-energy neutrino ever observed at $\sim$220~PeV, as a near-horizontal muon track underscores that the ultra-high-energy frontier is accessed through horizontal directions where the Earth's opacity above $\sim$100~TeV confines the observable sky to a narrow band around and above the horizon. Yet extending general-purpose detector architectures into this regime requires disproportionate increases in instrumentation, cost, and logistical complexity. A compelling alternative is to deploy specialized detectors that target this natural geometry. POLARIS (Pacific Ocean Large Area Radial Instrumented Sparse array) is a sparse planar deep-water Cherenkov array optimized for neutrino-induced muon tracks from horizontal directions in the multi-TeV to PeV regime. By rotating the conventional vertical string layout into a radial planar configuration, the detector presents maximal cross-section to horizontal tracks while naturally suppressing the down-going atmospheric background. With only 1100 optical modules, the five-arm design reaches point source and diffuse flux sensitivities at PeV energies competitive with detectors deploying several times more instrumentation. As a dedicated $\nu_\mu$ track detector, POLARIS provides the muon-flavor channel that tau-optimized experiments such as TAMBO and Trinity do not cover, enabling full flavor composition measurements from astrophysical sources. Using the Prometheus simulation framework, this study demonstrates that targeted sparse geometries can open new discovery space at the high-energy frontier at a fraction of the cost of general-purpose arrays.}

\keywords{neutrino astronomy, neutrino detectors, neutrino experiments}

\begin{document}
\maketitle
\flushbottom

\section{Introduction}
\label{sec:intro}

The era of neutrino astronomy has been established by a generation of cubic-kilometer-scale detectors. IceCube, operating at the South Pole since 2011, has demonstrated that a single neutrino telescope can sustain a remarkably versatile research program spanning astrophysics \cite{IceCube:2018cha, IceCube:2022der}, particle physics \cite{Ahlers:2018mkf}, and geophysics \cite{Donini:2018tsg}, while KM3NeT~\cite{KM3Net:2016zxf} and Baikal-GVD~\cite{Avrorin:2011zzc, Baikal-GVD:2020irv}, following the pioneering work of ANTARES~\cite{ANTARES:2011hfw}, are extending sky coverage from the Northern Hemisphere. These second-generation instruments employ either densely instrumented volume configurations or cluster-based geometries, as in GVD and as proposed for P-ONE~\cite{P-ONE:2020ljt}, targeting a broad energy range from GeV-scale atmospheric oscillations to PeV astrophysical sources.

The natural next step is to scale these proven architectures to larger volumes. IceCube-Gen2~\cite{IceCube-Gen2:2020qha, IceCube-Gen2:2024tdr} pursues this path at the South Pole, combining an extended optical array with a large surface detector and a radio component to cover energies from TeV to EeV at a single site. In the South China Sea, TRIDENT~\cite{TRIDENT:2022hql}, HUNT~\cite{Huang:2023mzt}, and NEON~\cite{Zhang:2024slv} aim to instrument multi-cubic-kilometer volumes with tens of thousands of optical modules, targeting broad all-flavor sensitivity. These efforts represent a substantial investment in neutrino infrastructure and, if realized, will significantly advance the field. However, general-purpose detectors that address the full breadth of the neutrino physics program must grow their instrumented volume and module count proportionally, with corresponding increases in cost and complexity. Given that neutrino observatories typically require two decades or more from design to mature physics output, replicating the general-purpose approach at ever-larger scale faces significant financial and organizational challenges.

A complementary strategy is to specialize. Rather than instrumenting ever-larger homogeneous volumes, one can exploit natural features of the detection environment to achieve targeted sensitivity at a fraction of the cost. The TAMBO project~\cite{Romero-Wolf:2020pzh,TAMBO:2025jio} exemplifies this philosophy: by deploying an established surface scintillator array in the deep valleys of the Peruvian Andes, it uses the mountain relief as a natural converter for tau neutrino detection in the 1--100~PeV range. Trinity~\cite{Otte:2019aaf,Bagheri:2025fxh} pursues a similar physics channel through Earth-skimming tau neutrino detection using imaging air-Cherenkov telescopes, a mature technique from gamma-ray astronomy. Both projects achieve competitive sensitivity with instrumentation budgets far below those of cubic-kilometer arrays.

The case for specialized detectors is reinforced by the geometry of neutrino detection at the highest energies. The Earth becomes opaque to neutrinos above $\sim 100$~TeV, so that no single site provides instantaneous full-sky coverage in this regime. The PLEnuM framework~\cite{Schumacher:2025qca} has quantified the gain from combining geographically distributed telescopes, demonstrating that a planetary network of neutrino observatories can increase detection rates by up to an order of magnitude and enable real-time transient coverage that no individual detector can achieve alone.

However, tau-neutrino-optimized experiments such as TAMBO and Trinity, while addressing a crucial detection channel, access only part of the available astrophysical information. Neutrino flavor composition encodes fundamental properties of the acceleration and production mechanisms~\cite{Ackermann:2022rqc}, and extracting the full \textit{messenger yield}, defined here as the combined directional, spectral, and flavor information accessible through all neutrino detection channels, requires complementary coverage of the muon neutrino track channel. Charged-current $\nu_\mu$ interactions produce through-going muon tracks that offer the best angular resolution among all event topologies and have proven most effective for source identification.

Near-horizontal muon tracks, in particular, have emerged as uniquely powerful messengers for the most energetic astrophysical events. The IceCube-170922A neutrino from the blazar TXS~0506+056~\cite{IceCube:2018cha, IceCube:2018dnn}, a near-horizontal muon track at $\sim$290~TeV that triggered the first neutrino--source association, and KM3-230213A~\cite{KM3NeT:2025npi}, a near-horizontal track at $\sim$220~PeV constituting the highest-energy neutrino ever detected, both demonstrate that the horizontal band is the primary discovery channel above 100~TeV. Since the locations of ultra-high-energy neutrino sources remain largely unknown, full-sky coverage is essential, and detectors at different latitudes contribute complementary horizon bands. Yet current detector geometries, with vertical string layouts optimized for upgoing track reconstruction, introduce systematic acceptance losses near the horizon~\cite{KM3Net:2016zxf}, precisely the angular regime where the astrophysical signal-to-background ratio is most favorable at the highest energies. As noted by ref.~\cite{KM3Net:2016zxf}, larger string spacings are beneficial for the detection of high-energy neutrinos, suggesting that sparse, extended geometries may be better suited to this regime than the dense instrumentation required for broad-band detectors.

We propose POLARIS (Pacific Ocean Large Area Radial Instrumented Sparse array), a sparse planar deep-water array optimized for neutrino-induced muon tracks arriving from horizontal and near-horizontal directions in the multi-TeV to PeV regime. By rotating the conventional vertical string geometry into a planar configuration, the detector presents maximal cross-section to near-horizontal tracks while naturally suppressing down-going cosmic-ray secondaries. The design achieves radical sparsification, deploying only 1100 optical modules compared to $\sim$4100 for KM3NeT-ARCA and $\sim$24,000 for TRIDENT, while concentrating sensitivity where horizontal tracks dominate the astrophysical signal. POLARIS is complementary to existing and proposed detectors: it covers a region of the messenger yield inaccessible to tau-specific experiments, extends the geographic reach of the emerging planetary neutrino network, and offers a modular architecture suitable for independent deployment at Pacific sites or as an extension of P-ONE~\cite{P-ONE:2020ljt}. The design targets conditions representative of the western Pacific, with a candidate location off Taiwan's east coast evaluated in this work.

\section{Detector Design}
\label{sec:detector}

\begin{figure}[htbp]
\centering
\includegraphics[width=0.99\textwidth]{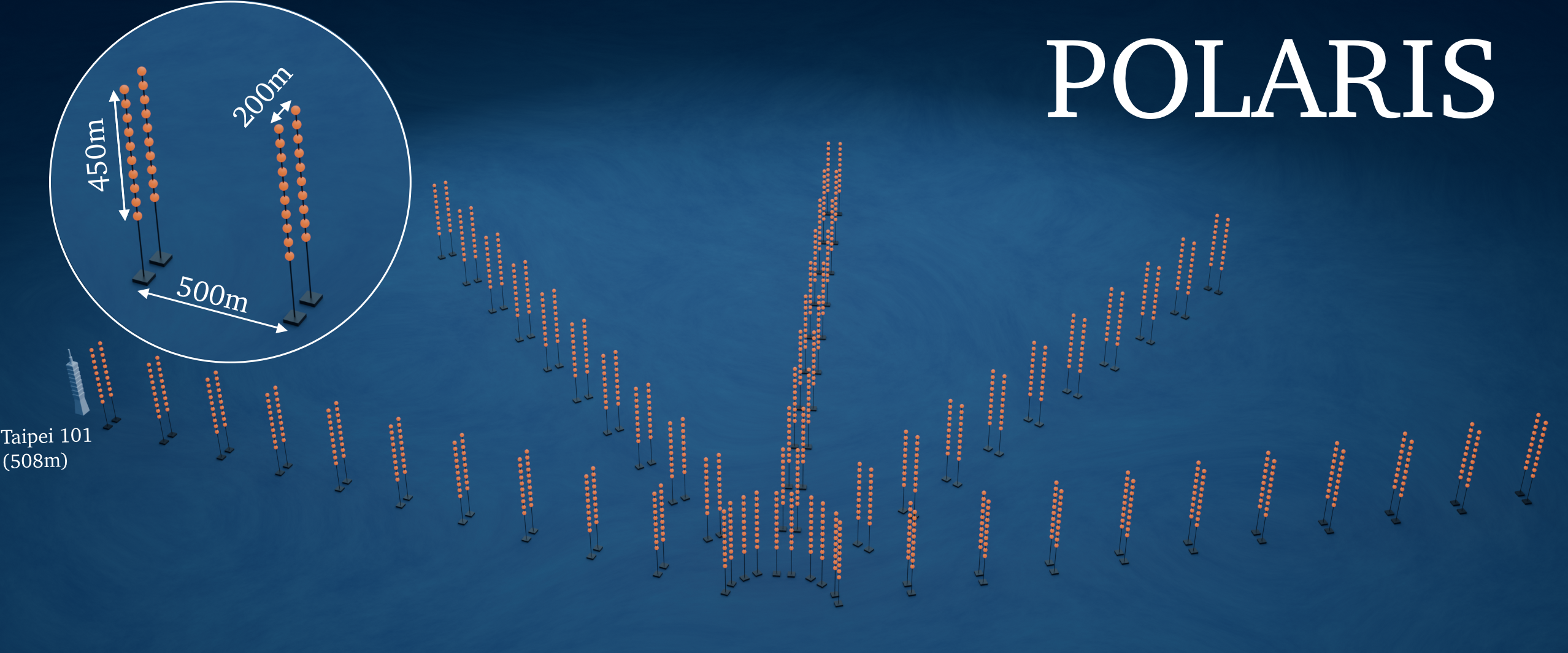}
\caption{Illustration of the five-arm POLARIS detector geometry with perpendicular string pairs (gates) at 500~m intervals. Each DOM in the top view represents a vertical string with 10~DOMs of 50~m spacing over 450~m height. Each arm is instrumented with 220~DOMs, totaling to 1100~DOMs for the whole detector.\label{fig:POLARIS}}
\end{figure}

The POLARIS design comprises five radial arms, each extending $5$~km from a central hub, arranged in a $120^{\circ}$ arc with $30^{\circ}$ angular spacing between adjacent arms, as shown in figure~\ref{fig:POLARIS}. We evaluate the detector geometry at a candidate site off Taiwan's east coast on a $3$~km flat plateau. Each arm deploys $11$ gates at $500$~m radial intervals, where a gate consists of two vertical strings separated by $200$~m in the transverse direction and instrumented with $10$~DOMs per string at $50$~m vertical spacing, spanning $450$~m of active height. The $500$~m gate spacing suppresses neutral-current cascade backgrounds by confining cascade-induced hits within a single gate, preventing the cross-gate coincidences that characterize PeV muon tracks. The $200$~m transverse string separation keeps the detector sparsely instrumented, reducing total DOM count while preserving the per-gate hit multiplicity required for track reconstruction. The five-arm configuration instruments $1100$ DOMs in total. A star-shaped configuration with three crossed arms separated by $60^{\circ}$ has also been evaluated and yields comparable effective area, but introduces non-uniform azimuthal acceptance whose impact on reconstruction performance requires dedicated study beyond the scope of this work.

\section{Event Selection and Topology Classification}
\label{sec:event_sel}

The sparse planar geometry implements natural topological discrimination between muon tracks and cascade events. Muon tracks traversing a radial arm produce sequential hits across spatially separated strings with monotonic timing progression, while cascade events deposit energy within several hundred meters, hitting at most two gates with near-simultaneous hits.

The trigger logic employs two-stage selection for muon tracks while removing cascade and atmospheric muon backgrounds: The local hit coincidence trigger for water detectors requires $\geq$2~PMT hits within 20~ns, suppressing K$^{40}$ radioactive backgrounds and noise ($\sim$kHz single-PMT rates \cite{Morton-Blake:2025vpp}), realized as 2 hits per DOM as no PMT settings can be adjusted in the Prometheus framework. The following geometric selection requiring $\geq 2$ hit DOMs at neighboring gates naturally eliminates cascades up to 1~PeV. The remaining PeV cascades are removed by a minimum total hit time spread across 1000~ns and a minimum hit separation length of 500~m (the maximum distance between any two causally connected PMT hits in the event). This geometric requirement selects through-going tracks along a detector arm while rejecting cascades before reconstruction, reducing computational overhead. This selection retains 40\% of signal tracks while rejecting $\sim$100\% of cascade backgrounds across the energy range from 10~TeV to 1~EeV. While the detector concept is designed for cascade event removal, no further optimization of the detector geometry was done.

\section{Astronomy Potential}
\label{sec:astro_potential}

\paragraph{Point source sensitivity.}
The point source sensitivity of the 5-arm POLARIS is derived by computing the minimum source flux required to yield a statistically significant detection above background in each energy bin, evaluated at the direction of KM3-230213A \cite{KM3NeT:2025npi} in the Southern Sky declination of $\delta = -7.8^{\circ}$. The variation across the visible sky is modest, consistent with a detector latitude of $\approx 23.99^{\circ}$N. The effective area $A_\mathrm{eff}(\cos\theta,\,\log_{10}E)$ is computed from Monte Carlo simulation and evaluated on the celestial sphere via a kernel density estimator, yielding a smooth two-dimensional map (see appendix~\ref{app:eff_area}). The signal acceptance within the search window is obtained by integrating this map over the point spread function at each energy.

Background contributions comprise the irreducible astrophysical neutrino flux, modeled using the IceCube-measured diffuse $\nu_\mu + \bar{\nu}_\mu$ spectrum \cite{Abbasi:2021qfz},
\begin{equation}
\label{eq:diffuse}
\phi^{\nu_\mu+\bar{\nu}_\mu} = 1.44\times10^{-18}\,\mathrm{GeV}^{-1}\mathrm{cm}^{-2}\mathrm{s}^{-1}\mathrm{sr}^{-1} \cdot \left(\frac{E}{100\,\mathrm{TeV}}\right)^{-2.37} \,,
\end{equation}
and the atmospheric neutrino contribution estimated from the data-driven model \textsc{daemonflux} \cite{Yanez:2023lsy}. The $5\sigma$ discovery flux is evaluated through a binned Poisson likelihood, the Cash statistic \cite{Cash:1979vz},
\begin{equation}
\label{eq:cash}
\mathcal{C} = 2\sum_i \left[\mu_i - n_i \ln(\mu_i)\right] \,,
\end{equation}
where $n_i$ and $\mu_i$ are the observed and expected counts in bin $i$. Only up-going and near-horizontal events with zenith angles $\theta \geq 80^{\circ}$ are retained, as shallower angles are dominated by the irreducible atmospheric muon flux. A point source detection is claimed at a test statistic value of $\text{TS} = 25$, corresponding to a $5\sigma$ discovery threshold.

The search window is centered on the source position and defined as the median angular resolution plus a $0.5^{\circ}$ margin. For POLARIS, we adopt an energy-dependent median angular resolution of IceCube \cite{IceCube:2018ndw} as a conservative benchmark: although POLARIS operates at larger instrumented volume, its sparser DOM density offsets this advantage, and IceCube's published resolution carries no error bands, making its adopted value a lower bound rather than a central estimate. At few PeV, this angular resolution reaches $\sim 0.3^{\circ}$. The $0.5^{\circ}$ margin reflects the angular resolution uncertainty reported by TRIDENT \cite{Morton-Blake:2025vpp}.

\begin{figure}[htbp]
\centering
\includegraphics[width=\textwidth]{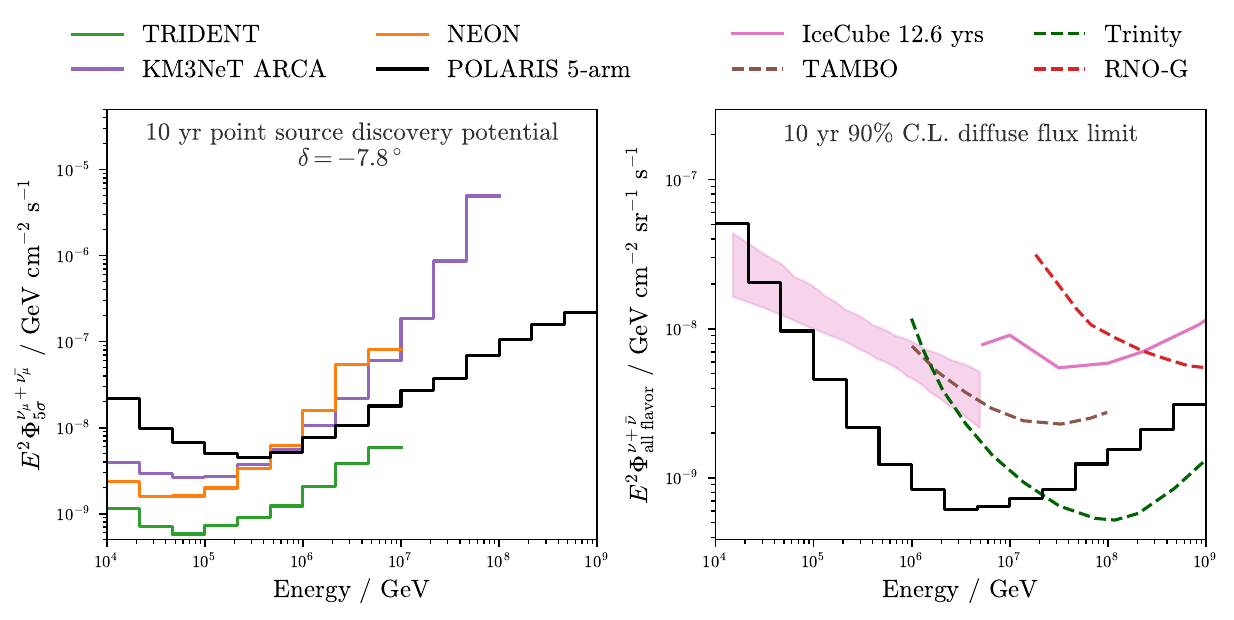}
\caption{\textit{Left:} The $\nu_\mu + \bar{\nu}_\mu$ flux required for $5\sigma$ point source detection at $\sin\delta = 0$ for ten years of operation per energy bin for 5-arm POLARIS. Sensitivity curves for IceCube, KM3NeT ARCA, NEON and TRIDENT are computed using their published effective areas and angular resolutions following the identical analysis procedure as for POLARIS. \textit{Right:} Differential $90\%$ C.L.\ sensitivities to an all-flavor diffuse neutrino flux for ten years of operation for POLARIS compared to current and proposed experiments. The pink line shows the IceCube experimental upper limits for 12.6 years of operation \cite{IceCubeCollaborationSS:2025jbi}. The diffuse measurement of the $\nu_{\mu}$ with 9.5 years of IceCube data is depicted as pink band \cite{Abbasi:2021qfz}. Dashed lines indicate expected 10-year sensitivities for TAMBO (22{,}000 detectors) \cite{Romero-Wolf:2020pzh,Ackermann:2022rqc}, Trinity (18 telescopes) \cite{Otte:2019aaf,Ackermann:2022rqc}, RNO-G \cite{RNO-G:2020rmc,RNO-G:2021hfx,Ackermann:2022rqc}.\label{fig:sensitivity}}
\end{figure}

To enable a uniform comparison, the sensitivity curves for IceCube, KM3NeT ARCA, NEON, and TRIDENT are recomputed following the identical procedure, using their published effective areas \cite{IceCube:2021xar,KM3NeT:2024paj, Zhang:2024slv,TRIDENT:2022hql} and angular resolutions \cite{IceCube:2018ndw,KM3NeT:2024paj,Zhang:2024slv,Morton-Blake:2025vpp}. The comparison is restricted to Northern Hemisphere detectors with publicly available effective areas. The resulting ten-year $5\sigma$ discovery flux is shown in the left panel of figure~\ref{fig:sensitivity}. TRIDENT achieves a superior discovery flux across the energy range shown due to its larger effective area; the TRIDENT curve is truncated at $10\,\mathrm{PeV}$, the upper boundary of their published effective area \cite{Morton-Blake:2025vpp}. POLARIS outperforms KM3NeT-ARCA and NEON above $\sim$500~TeV, with an order-of-magnitude improvement above $10\,\mathrm{PeV}$, reflecting the design optimization toward the PeV regime. The NEON effective area is derived from a less restrictive event selection than adopted here, making this comparison conservative for POLARIS. Adopting the angular resolution of TRIDENT instead of IceCube would improve the POLARIS point source sensitivity by ${\sim}26.3\%$ at energies of a few PeV.

\paragraph{Diffuse flux sensitivity.}
The diffuse flux analysis shares the likelihood framework and background model of the point source analysis but differs in sky coverage and signal integration. Rather than restricting to a search window around a candidate source position, the full sky is divided into ten bins of $\cos\theta$ to exploit the distinct zenith-dependent profiles of the astrophysical signal and atmospheric backgrounds. For each $(\cos\theta,\,\log_{10}E)$ bin, the expected signal is computed by folding $\Phi(E)$ through $A_\mathrm{eff}(\cos\theta,\,\log_{10}E)$, and the $90\%$ C.L.\ upper limit per energy bin is defined as the flux normalization for which $\mathcal{C} = 2.71$. Because the analysis uses $\nu_\mu + \bar{\nu}_\mu$ only, the resulting limit is scaled by a factor of three under the assumption of equal flavor composition at Earth, enabling direct comparison with all-flavor upper limits from other experiments.

The differential $90\%$ C.L.\ sensitivity for ten years of operation is shown in the right panel of figure~\ref{fig:sensitivity}. POLARIS reaches below the published limits of IceCube and KM3NeT across the fitted energy range, crossing below the IceCube diffuse muon-neutrino flux measurement above $20\,\mathrm{TeV}$. In the PeV regime, where POLARIS is optimized, the projected sensitivity surpasses TAMBO at all energies, Trinity up to ${\sim}20\,\mathrm{PeV}$, and RNO-G up to ${\sim}1\,\mathrm{EeV}$.

\section{Discussion}
\label{sec:discussion}

The point source and diffuse flux results both reflect the core geometric trade-off of the POLARIS design. The 500~m gate spacing that isolates PeV muon tracks from cascade backgrounds sets a practical lower energy threshold near 10~TeV, below which multi-gate coincidence probability falls and angular resolution degrades. Sources dominated by sub-TeV emission, including the Galactic Center and the Galactic plane diffuse flux measured by IceCube~\cite{HESS:2016pst, Celli:2016uon}, lie outside the POLARIS sensitivity window. This threshold is inherent to the sparse geometry and represents a deliberate trade-off: sensitivity to horizontal high-energy tracks at the expense of broad-band energy coverage.

Above 200~TeV, POLARIS surpasses KM3NeT-ARCA in point source sensitivity and outperforms P-ONE at PeV energies despite a comparable optical module count~\cite{Twagirayezu:2023cpv}, using 74\% and 94\% fewer modules than KM3NeT-ARCA and TRIDENT respectively. TRIDENT maintains superior point source sensitivity across the full energy range, consistent with its much larger instrumented volume. The all-flavor diffuse flux sensitivity, derived from the $\nu_\mu$ channel assuming equal flavor composition at Earth, crosses below the IceCube diffuse muon-neutrino flux measurement above 20~TeV and exceeds the $\nu_\tau$ sensitivities of TAMBO at all energies and Trinity up to ${\sim}20$~PeV, as well as RNO-G up to ${\sim}1$~EeV. Together with $\nu_\tau$-sensitive instruments such as TAMBO and Trinity, a network of specialized detectors at different latitudes would cover the messenger yield across flavors and sky regions, as argued in the PLEnuM framework~\cite{Schumacher:2025qca}. At Taiwan's latitude ($\approx 24^{\circ}$N), POLARIS would contribute a horizon band distinct from those of IceCube, KM3NeT, and Baikal-GVD. Even if IceCube-Gen2 and the proposed large-volume arrays in the South China Sea are fully realized, geographically distributed specialized detectors would remain an irreplaceable source of real-time alerts, since no single site can monitor the full high-energy neutrino sky at any given time.

The radial arm architecture is intrinsically scalable. Adding arms increases effective volume and azimuthal acceptance, while reducing their number lowers cost proportionally, allowing the detector to be sized to available resources without redesigning the basic instrumentation unit. This scalability, combined with the significantly lower module count compared to general-purpose arrays, reduces the investment required to enter neutrino astronomy and could enable nations in the Asia-Pacific region that do not yet operate or contribute to a neutrino telescope to participate in the field.

Several aspects of the present study require further development. Site-specific optical properties strongly influence effective area~\cite{Morton-Blake:2025vpp}, and dedicated measurements at the proposed Taiwan site are needed to refine the sensitivity projections. The event selection achieves complete cascade rejection but retains only 40\% of signal tracks; graph neural network reconstruction and event selection~\cite{Orsoe:2025aqu} is currently under development and is expected to improve both signal retention and angular resolution. Adopting the angular resolution reported by TRIDENT rather than the conservative IceCube benchmark used here would improve point source sensitivity by ${\sim}26.3\%$ at energies of a few PeV. Extension of the detector concept to tau double-bang topologies and cascade measurements at the highest energies, as well as the development of deployment techniques suited to the sparse radial geometry, will be addressed in future work.

\acknowledgments

We thank Marie-C\'ecile Piro, Juan Pablo Ya\~nez, and Dvir Hilu for fruitful discussions at the initial stage of the detector design. We also thank Iwan Morton-Blake, Akimichi Taketa, Hiroyuki Sagawa, and Jeffrey Lazar for helpful discussions. We acknowledge support from Academia Sinica (Grant No.~AS-GCS-113-M04) and the National Science and Technology Council (Grant No.~113-2112-M-001-060-MY3). KH acknowledges support from the Postdoctoral Scholar Program of the Academia Sinica (PD-11401-M-3854). The authors acknowledge the use of computational resources provided by the Academia Sinica Grid Computing Center (ASGC), a Core Facility of Academia Sinica (AS-CFII-114) and an NSTC Core Computing Center funded by the National Science and Technology Council (NSTC 113-2740-M). Claude by Anthropic has been used for language and code improvements.

\appendix

\section{Simulation Framework}
\label{sec:sim}

Detector performance is characterized through Monte Carlo simulations using the \texttt{Prometheus} framework \cite{Lazar:2023rol}, a modular neutrino telescope simulation toolkit supporting configurable detector geometries. Neutrino interactions are generated with \texttt{LeptonInjector} \cite{IceCube:2020tcq} over the energy range 10~TeV to 1~EeV, with separate simulation sets for $\nu_e$, $\nu_\mu$, and $\nu_\tau$ and their antiparticle counterparts, covering both charged-current and neutral-current interactions for each flavor. Events are injected following an unbroken $E^{-1}$ power-law spectrum in ranged mode with uniform sampling in column density and a cosine-uniform angular distribution \cite{Zhu:2024ubz}. For computational efficiency, simulations are performed for a single radial arm; effective areas are scaled by a factor of five to represent the full five-arm detector. The arm separation is large enough that cross-arm coincidences are not expected, so the scaling assumes statistically independent arm responses and yields a lower bound on the effective area.

The Earth model follows PREM \cite{Dziewonski:1981xy}, modified to include a 3~km water column above the solid Earth and below the atmosphere, consistent with the depth convention adopted by in-water arrays such as for ARCA in \texttt{Prometheus} \cite{Lazar:2023rol}. Detector modules are positioned between depths of 2475~m and 2925~m below the sea surface. Within the \texttt{Prometheus} pipeline, charged lepton propagation is handled by \texttt{PROPOSAL} \cite{Koehne:2013gpa}, which accounts for continuous ionization losses and stochastic processes including bremsstrahlung, pair production, and photonuclear interactions. Cherenkov light yields are then computed by \texttt{Fennel} \cite{fennel2022_github} using parameterizations derived from \textsc{Geant4} \cite{GEANT4:2002zbu,Radel:2012kw} shower simulations. Photon propagation in the water medium is handled by \texttt{Hyperion} via ray-tracing \cite{Lazar:2023rol}. Event weights mapping the injected sample to physical neutrino fluxes are computed with LeptonWeighter \cite{IceCube:2020tcq}.

Optical properties are set to the TRIDENT South China Sea measurements \cite{TRIDENT:2022hql}, with absorption length $\lambda_{\rm abs}\simeq 27$~m and scattering length $\lambda_{\rm sca}\simeq 63$~m at 450~nm; the South China Sea site is adopted due to its geographic proximity to Taiwan and the absence of site-specific optical measurements at the proposed deployment location. Water chemical composition is set to the P-ONE defaults provided in \texttt{Prometheus} \cite{Lazar:2023rol,Bailly:2021dxn}. Photomultiplier quantum efficiency is set to 0.2 and noise rate to $10^3$~Hz per module, following the baseline values defined in the \texttt{Prometheus} framework \cite{Lazar:2023rol}.

\section{Effective Area}
\label{app:eff_area}

\begin{figure}[htbp]
\centering
\includegraphics[width=0.6\textwidth]{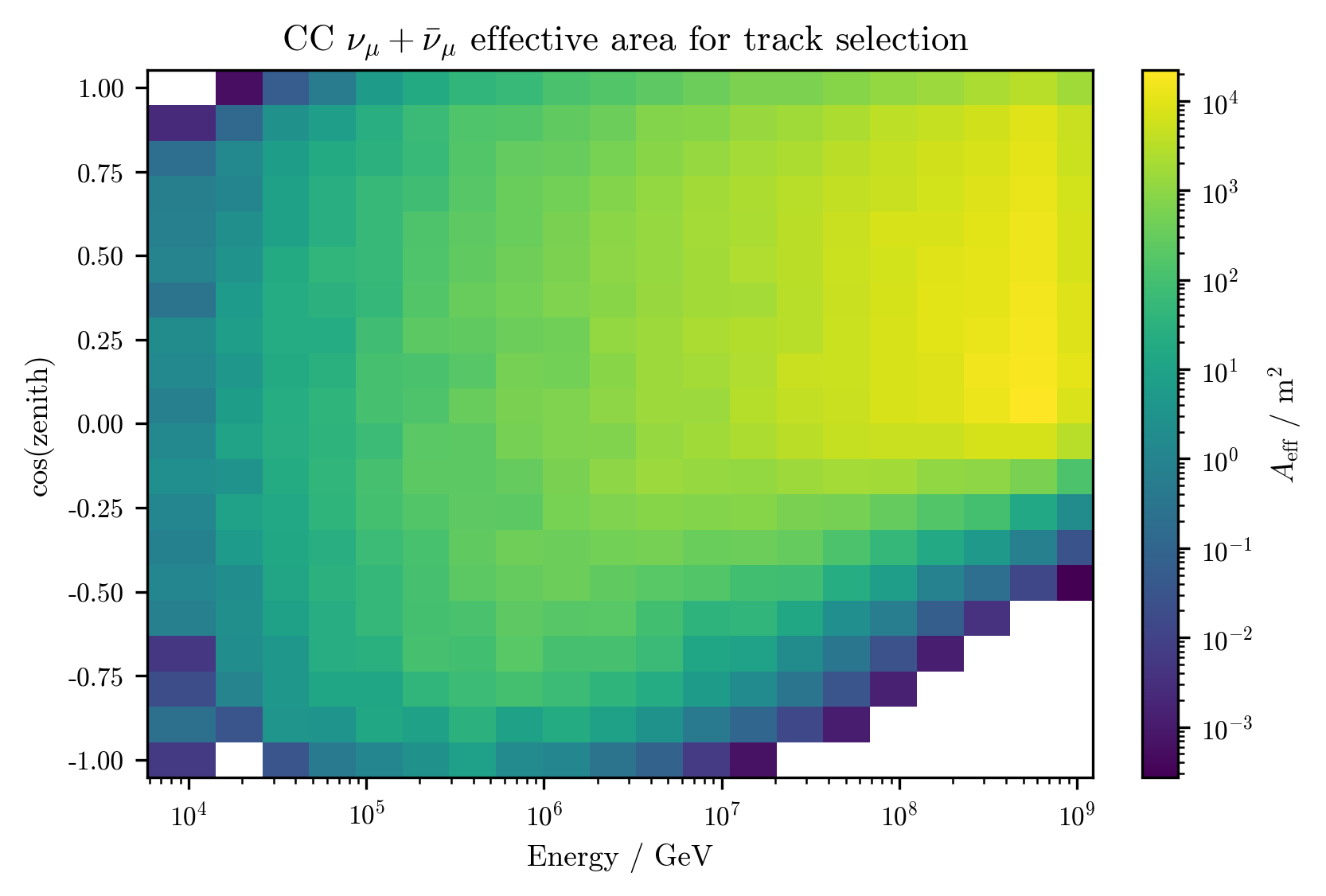}
\caption{Effective area for CC muon and anti muon neutrinos for the track selection of the 5-arm POLARIS geometry.\label{fig:eff_area}}
\end{figure}

The effective area for charged-current muon neutrinos is shown in figure~\ref{fig:eff_area}. The effective area from LeptonInjector is corrected for Earth absorption probability with nuSQuiDS \cite{Arguelles:2021twb}.

\bibliographystyle{JHEP}
\bibliography{references}

\end{document}